# Coupling Power Laws Offers a Powerful Method for Problems such as Biodiversity and COVID-19 Fatality Predictions

Zhanshan (Sam) Ma
[1]Computational Biology and Medical Ecology Lab
Kunming Institute of Zoology
Chinese Academy of Sciences
Kunming 650222, China
[2]Center for Excellence in Animal Genetics and Evolution
Chinese Academy of Sciences
Kunming 650222, China
ma@vandals.uidaho.edu

## Abstract

Power laws have been found to describe a wide variety of natural (physical, biological, astronomic, meteorological, geological) and man-made (social, financial, computational) phenomena over a wide range of magnitudes, although their underlying mechanisms are not always clear. In statistics, power law distribution is often found to fit data exceptionally well when the normal (Gaussian) distribution fails. Nevertheless, predicting power law phenomena is notoriously difficult because some of its idiosyncratic properties such as lack of well-defined average value, and potentially unbounded variance. TPL (Taylor's power law), a power law first discovered to characterize the spatial and/or temporal distribution of biological populations and recently extended to describe the spatiotemporal heterogeneities (distributions) of human microbiomes and other natural and artificial systems such as fitness distribution in computational (artificial) intelligence. The power law with exponential cutoff (PLEC) is a variant of power-law function that tapers off the exponential growth of power-law function ultimately and can be particularly useful for certain predictive problems such as biodiversity estimation and turning-point prediction for COVID-19 infection/fatality. Here, we propose coupling (integration) of TPL and PLEC to offer improved prediction quality of certain power-law phenomena. The coupling takes advantages of variance prediction using TPL and the asymptote estimation using PLEC and delivers confidence interval for the asymptote. We demonstrate the integrated approach to the estimation of potential (dark) biodiversity and turning point of COVID-19 fatality. We expect this integrative approach should have wide applications given the duel relationship between power law and normal statistical distributions.

# Introduction

Taylor's power law (TPL), first discovered by entomologist and ecologist L. R. Taylor (1961) in investigating the spatial distribution of insect populations more than a half century ago (Taylor & Taylor 1977, Taylor *et al* 1983, 1988, Taylor 1984, 2018, 2019), has been expanded far beyond its original domains of agricultural entomology and population ecology (Eisler *et al*. 2008, Fronczak & Fronczak 2010, Taylor 2019, Ma 1991, 2012, 2013, Ma & Taylor 2020). The TPL is one form of power laws that describe the distributions of a wide variety of natural and man-made phenomena over a wide range of scales (Cohen *et al*. 2012, Cohen & Xu 2015, Reuman et al 2017). Power law patterns have been discovered/rediscovered in astronomy, biology, computer science, ecology, criminology, economics, finance, geology, mathematics, meteorology, physics, statistics, and especially in inter-disciplinary fields (Stumpf & Porter 2012, Eliazar 2020).

A power law describes a non-linear functional relationship between two variables—one varies as a power of another [*e.g.*, $f(x) = a(x)^b$] and has certain properties including scale invariance, lack of well-defined average value, and universality (Eisler et al. 2008, Fronczak & Fronczak 2010, Eliazar 2020, Stumpf & Porter 2012). The scale invariance is exhibited by a simple log-transformation of power law into a straight-line (linear) on log-log scale {*e.g.*, $\ln[f(x)] = \ln(a) + b\ln(x)$}, and it also specifies that all power laws with a particular scaling exponent are equivalent up to constant factors, *e.g.*, $f(cx) = a(cx)^b = c^b f(x) \propto f(x)$. The lack of well-defined average value refers to a reality that arithmetic mean or average is a poor indicator for majority of the power-law variables (*e.g.*, the average income of a population including a billionaire). A power law usually has a well-defined mean *only* for certain range of its scaling exponents, and the variance of power law seems disproportionally large and is frequently not well-defined, which explains the association between power law phenomena and black swan behavior. This also makes many classic statistical methods that are based on normal distribution or based on the homogeneity of variance in applicable to data of power law phenomena. A third property of power law is the universality that is to do with the scale invariance or the equivalence of power laws with a particular scaling exponent. In dynamic systems, diverse systems with the same power-law scaling exponents (also known as critical exponents) can exhibit identical scaling behavior and share the same fundamental dynamics as they approach criticality such as phase transitions. Systems with the same critical exponents are classified as belonging to the same universality class (Eisler *et al*. 2008, Fronczak & Fronczak 2010, Stumpf & Porter 2012, Taylor 2019, Ma & Taylor 2020, Eliazar 2020).

TPL, as one of the most well-known power laws in ecology and biology, share the three general properties of power laws mentioned above. It differs with other power laws in choosing its two variables: the mean and variance of population abundances (counts) (Taylor 1961, 1984, 1986, 2019), *i.e.*, $V = aM^b$. It has also been rediscovered in many other fields beyond its original domain of population ecology such as epidemiology, genomics and metagenomics, and computer science (Cohen *et al*. 2012, Cohen & Xu 2015, Reuman et al 2017, Taylor 2019, Ma 2012, 2013, 2015, Ma & Taylor 2020). It was extended to community ecology, especially the community and landscape ecology of human microbiomes (Ma 2015, 2021, Ma & Taylor 2020). Compared with other power laws, TPL has two somewhat unique characteristics. The first is that its scaling parameter (exponent) measures the population (community) spatial heterogeneity or temporal stability. This has to do with the fact that the variance to mean ratio (V/M) is a measure of dispersion of data points (population abundances or counts), while dispersion, aggregation and heterogeneity essentially characterize the same or similar system property (Ma & Taylor 2020, Ma 2020b). Indeed, TPL scaling parameter (*b*) can be used to measure heterogeneity at population, community and landscape levels, respectively, depending on the level the TPL model was built. The second characteristic of TPL is also related to the variance and mean: the relationship can be utilized for designing sampling schemes since the variance (level of variation or heterogeneity) determine the sampling efforts (sample sizes) necessary for estimating the population (species) abundances reliably (*e.g.*, Taylor 2018, Ma 2020a). We take the advantages of TPL in this study to improve the quality of prediction/estimation because variance or standard deviation is the foundation for computing confidence interval of estimation.

Species-area relationship (SAR) is another classic power law in ecology, which related the number of species (species richness: *S*) and the area (A) of species habitat, in the form of $S=cA^z$. Ma (2018a, 2019) further extended the SAR to general diversity-area relationship (DAR) by replacing species number (richness) with the general diversity measured in Hill numbers. Ma (2018a, 2018b, 2019) further introduced PLEC (power law with exponential cutoff) model to describe DAR and proposed the concept of maximal accrual diversity (MAD). Based on PLEC model for DAR, he derived the estimation of MAD. MAD can be considered as proxy of potential or dark diversity, which refers to diversity that includes both local diversity and the portion of diversity that are absent locally but present regionally (or in regional species pools). In other words, potential diversity measures both visible and invisible (dark) diversities, and is of obvious significance for biodiversity conservation. Similar to SAR/DAR, there is so-called STR (species-time relationship) or DTR (diversity-time relationship) (Ma 2019). The PLEC version of DTR was successfully applied to predict the inflection points (tipping points) of COVID-19 infections (Ma 2020c).

PLEC, as a variant of power law, has more general applications beyond the above-mentioned SAR/DAR/STR/DTR/COVID-19 predictions (Ma 2018a, 2018b, 2019, 2020c). PL behaves (grows or declines) exponentially, especially at late stages, and the PLEC possesses an exponential-cutoff parameter that ultimately taper off the unlimited growth or decline ultimately. Therefore, PLEC model is of important practical significance when prediction or estimation is needed. However, existing PLEC modeling can only provide point estimation, not the interval of estimation.

The present article is aimed to integrate the TPL with PLEC model with the objective to improve the predictive power of PLEC model. Specifically, by harnessing the capacity of TPL in estimating the variance (standard deviation), we develop an approach to offering confidence intervals for the estimation of PLEC quantities (see Fig 1). We demonstrate our method with the estimations of potential gut microbiome diversity and COVID-19 fatalities.

## Material and Methods

**Taylor's power law (TPL)**

TPL is one form of power laws, and it establishes the relationship between the variance and mean of a random variable $Y$ (*e.g.*, population counts or abundances of biological populations) as a power function:

$$Var(Y) = V = aM^b \quad (1)$$

where $V$ & $M$ are the variance and mean of random variable $Y$; $a$ & $b$ are the parameters that can be estimated by fitting TPL to a series of spatial or temporal samples of populations. TPL can be fitted by a simple log-transformation, which generates:

$$\ln(V) = \ln(a) + b\ln(M) \quad (2)$$

Ma (2015) extended TPL to community level by specifying $Y$ as species abundance, $M$ as the mean species abundance (size) *per* species in a community, and $V$ is the corresponding variance. By regressing $V$-$M$ across a series of communities (samples), one obtains type-I TPLE (TPL extension) for community spatial heterogeneity and type-II TPLE for community temporal stability. Similarly, there were type-III for mixed-species spatial heterogeneity and type-IV for mixed species temporal stability. The four TPLEs have the exactly same mathematical form as the original TPL (1)-(2), but the variables and parameters are defined and interpreted differently.

Taylor (2019) conjectured that TPL only applied to integers such as population counts (abundances), and it works poorly for ratios and very poorly for bounded ratios.

In this study, we take the advantages of TPL/TPLEs to estimate variance ($V$) corresponding to mean ($M$). The variance or its squared root (standard deviation) provides necessary quantities for estimating confidence intervals of PL or PLEC models.

**Power law with exponential cutoff (PLEC) model**

PLEC is a variant of power law (PL) model, and it was initially used to extend another classic power law in ecology, *i.e.,* the species-area relationship (SAR) (Watson 1835, Preston 1960). The PL model for SAR (species-area relationship) is:

$$S = cA^z \qquad (3)$$

where $S$ is the number of species and $A$ is the area of habitat occupied by $S$ species.

Ma (2018a) extended SAR to general DAR (diversity-area relationship) by replacing the species richness (number of species) with general diversity (in Hill numbers).

$$^qD = cA^z \qquad (4)$$

where $^qD$ is diversity measured in the *q-th* order Hill numbers, $A$ is *area*, and $c$ & $z$ are parameters.

The PLEC model for DAR is:

$$^qD = cA^z \exp(dA), \qquad (5)$$

where $d$ is a third parameter (taper-off parameter) and should be negative in DAR scaling models, and $\exp(dA)$ is the exponential decay term that eventually overwhelms the power law behavior at very large value of $A$. The PLEC was originally introduced to SAR modeling by Plotkin et al. (2000) and Ulrich & Buszko (2003) (also *see* Tjørve 2009), and Ma (2018a) extended it to DAR.

Ma (2018a) further derived the asymptote of PLEC model, and termed it as the maximal accrual diversity or potential diversity.

$$A_{max} = -z/d \qquad (5)$$

$^qD$ may have a maximum in the following form:

$$Max(^qD) = c(-\frac{z}{d})^z \exp(-z) = cA_{max}^z \exp(-z) \qquad (6)$$

There are similar species-time relationship (STR) and corresponding diversity-time relationship (DTR) (Preston 1960, Ma 2018b). STR/DTR has the exactly same PL/PLEC models as SAR/DAR described previously, but the data used to fit the models are different and so do the model parameters (Ma 2018b).

Ma (2020c) adapted STR/DTR model to predict the inflection points of COVID-19, in which maximal accrual or potential diversity is equivalent to maximal infection numbers. In STR/DTR modeling, a convention is to use parameter $w$ in place of the $z$ of SAR/DAR as diversity-time scaling parameter.

In the present study, we use the PLEC-DAR model to demonstrate the prediction of gut microbiome diversity, and PLEC-DTR model to demonstrate the prediction of COVID-19 fatalities, both augmented by TPL model to get their confident intervals, as outline below:

**Coupling TPL and PLEC models for predicting the interval of COVID-19 fatalities**
**Step (*i*)** Use PLEC model (eqn. 5), adapted for fitting the FTR (fatality-time relationship) datasets as follows, *i.e.*,

$$F = cT^w \exp(dT), \qquad (7)$$

where $T$ is the time in day, and $F$ is the fatality, $c$, $w$ and $d$ are PLEC-FTR parameters. The taper-off effects of parameter $d$ is usually rather weak before the fatality numbers reach peak, it is reasonable to treat $z$ as an approximation to the *fatality growth rate*, and $c$ as an approximation to the *initial fatality number*. To fit PLEC-FTR model (eqn. 7), we adopted a nonlinear optimization algorithm implemented as an R function "nlsLM" in R package "minpack.lm" (https://www.rdocumentation.org/packages/minpack.lm/versions/1.2-1/topics/nlsLM). Since $T_{max}>0$ is a necessary condition for the PLEC model to be biomedically sound, a constraint $d<0$ was imposed for the non-linear fitting of the PLEC-FTR model.

**Step (*ii*)** Compute maximal accrual fatality number (MAF) using eqn. (5) & (6), adapted as:

$$F_{max} = c(-\frac{w}{d})^w \exp(-w) = cT_{max}^w \exp(-w) \qquad (8)$$

$$T_{max} = -w/d \qquad (w>0, d<0) \qquad (9)$$

**Step (*iii*)** Use TPL model (eqn. 1) for fitting the spatiotemporal aggregation (heterogeneity) of fatality numbers, *i.e.*, adapting the original TPL (eqn. 1) as the following TPL for fatality aggregation:

$$V = a\overline{F}^b \qquad (10)$$

where $\overline{F}$ is the mean fatality number of COVID-19 and $V$ is the corresponding variance; $a$ & $b$ are the parameters. Parameters $a$ & $b$ are estimated by fitting eqn. (10) to spatiotemporal data of COVID-19 fatality, using the same scheme/procedures as used for fitting TPL to COVID-19 infection numbers (Ma 2020).

**Step (*iv*)** Compute the variance ($V$) and standard deviation ($\sqrt{V}$) based on eqn. (10) for fatality ($F$) (eqn. 7) or maximal accrual fatality ($F_{max}$) (eqn. 8).

**Step (*v*)** Compute the lower and upper limits of 95% confidence interval of COVID-19 fatality with the following pair of equations:

$$lower = F - 1.96 \times \sqrt{V/n} \qquad (11a)$$

$$lower = F_{max} - 1.96 \times \sqrt{V_{max}/n} \qquad (11b)$$

$$upper = F + 1.96 \times \sqrt{V/n} \qquad (12a)$$

$$upper = F_{max} + 1.96 \times \sqrt{V_{max}/n} \qquad (12b)$$

where $n$ is the number of time points that correspond to $F$ or $F_{max}$ in (eqn. 10).

With eqns.(11a) and (12a), one can obtain the confidence interval of COVID-19 fatalities at any time (day) points; alternatively, with eqns.(11b) and (12b), one can obtain the confidence interval of maximal accrual of COVID-19 fatality.

When $F_{max}$ cannot be predicted (too early to predict), the PL (power law) model for FTR can be used to complete the above procedures for estimating the intervals of $F$, i.e., by setting $d=0$, there is PL model for $F = cT^w \exp(dT) = cT^w$.

**Coupling TPL and PLEC models for predicting the gut microbiome diversity**

Similar to the previous integration of TPL and PLEC for estimating the confidence intervals of COVID-19 fatalities, here we specify the procedures for predicting the confidence intervals of gut microbiome diversity.

**Step (*i*)** Use PLEC model (eqn. 5) for fitting the DAR (diversity-area relationship) datasets, *i.e.*,

$$^qD = cA^z \exp(dA), \qquad (13)$$

where $A$ is the number of individuals, and $^qD$ is the diversity in Hill numbers, $c, z$ and $d$ are PLEC-DAR parameters. To fit the PLEC-DAR model, we use the same non-linear optimization procedures as described previously.

**Step (*ii*)** Compute maximal accrual diversity number (MAD) using eqn. (5) & (6).

**Step (*iii*)** Use TPL model (eqn. 1) for fitting the mean diversity and variance relationship:

$$V = a\overline{D}^b \qquad (14)$$

where $\overline{D}$ is the mean diversity (Hill numbers) of COVID-19 and $V$ is the corresponding variance; $a$ & $b$ are the parameters. Parameters $a$ & $b$ are estimated by fitting eqn. (10) to spatiotemporal data of Hill numbers, using the same scheme/procedures as described above for COVID-19 fatality.

**Step (iv)** Compute the variance ($V$) and standard deviation ($\sqrt{V}$) based on eqn. (10) for diversity ($D$) (eqn. 5) or maximal accrual diversity ($D_{\max}$) (eqn. 6).

**Step (v)** Compute the lower and upper limits of 95% confidence interval of diversity with the following pair of equations:

$$lower = D - 1.96 \times \sqrt{V/n} \qquad (15a)$$

$$lower = D_{\max} - 1.96 \times \sqrt{V_{\max}/n} \qquad (15b)$$

$$upper = D + 1.96 \times \sqrt{V/n} \qquad (16a)$$

$$upper = D_{\max} + 1.96 \times \sqrt{V_{\max}/n} \qquad (16b)$$

where $n$ is the number of samples corresponding to $D$ or $D_{\max}$ (eqn. 14). With eqns.(15a) and (16a), one can obtain the confidence interval of diversity at any diversity accrual points; alternatively, with eqns.(16a) and (16b), one can obtain the confidence interval of maximal accrual of diversity in Hill numbers.

When $D_{\max}$ cannot be predicted (too early to predict), the PL model for DAR can be used to complete the above procedures for estimating the intervals of $D$, i.e., by setting $d=0$, there is PL model for $D = cT^z \exp(dA) = cA^z$.

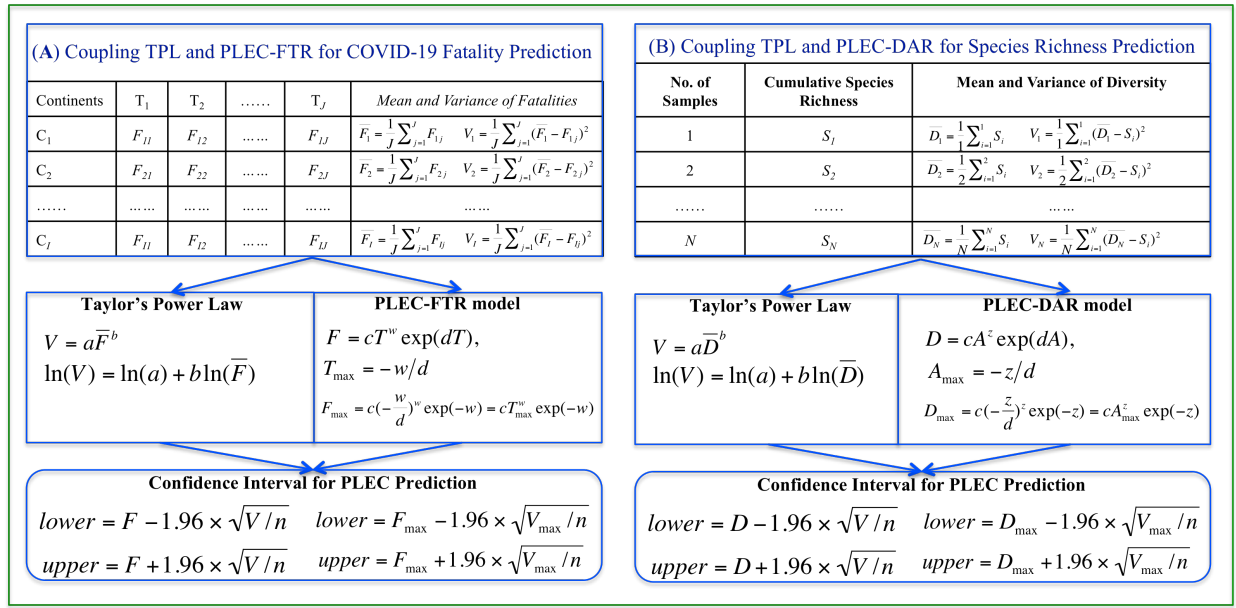

**Fig 1**. A diagram illustrating the coupling of TPL and PLEC models: for predicting COVID-19 fatality (A: the left block) and gut microbiome diversity (B: the right block)

## Results

**Coupling TPL and PLEC-FTR for predicting the intervals of COVID-19 fatalities**

The worldwide COVID-19 fatality numbers are available from the following website (https://github.com/CSSEGISandData/COVID-19) managed by John Hopkins University. Since the objective of this study is to demonstrate the feasibility of the coupling power law approach, we only extracted continent level data for demonstrative purpose. For the country-level predictions, which are too extensive to cover in this article, we have another separate report.

Fig 1A illustrated the procedures to predict COVID-19 fatality and Table 1 listed the predictions for six continents and the whole world. The PLEC modeling succeeded in all continents and the world, except for the Asia. The failure in Asia should be that the new wave of outbreak in India is still too early to foresee the fatality turning point, as discussed in Ma (2020) for the similar prediction of COVID-19 infections.

In Table 1, the first five columns are self-evident given they are simply the PLEC-FTR parameters. The next three columns are the predictions by the PLEC model, the maximal accrual fatality (number) ($F_{max}$) and the days ($T_{max}$) (Julian days or Calendar date) at which $F_{max}$ occurs. The next column is the actual fatality numbers at May 21[st], which happened to be the date we had completed the modeling work of this study, and which was listed to allow for a quick and rough reality-check. The next column is the "completion level"—the percentage of past fatality over maximal accrual fatality ($F_{max}$). The last two columns are the novel contribution of this study, *i.e.*, the lower and upper limits of predicted fatality numbers, which are not possible without the coupling of both the power laws (TPL and PLEC-FTR models).

Table 2 listed the fatality prediction for the Asia based on the PL-FTR model, for which the PLEC model failed. The predictions of PL model should be treated with caution, and are only of rough reference value. As explained previously, when the PLEC-FTR modeling efforts fails, it is usually that the outbreak is still in early stage and there is not yet sufficiently long time-series datasets to allow for the fitting of PLEC model. Although PL-FTR model can be fitted in these cases, the predictions from the PL model are not sufficiently reliable.

Similar to the predictions of COVID-19 infections, there are some standard pre-processing procedures to take before fitting the PLEC-FTR to the fatality-time (day) datasets. For example, proper selection of starting point by truncating early data points (possibly including whole previous pandemic waves) could be necessary for successful model fitting. In fact, the fitting results presented in Table 1 were obtained by setting the starting date for modeling on March 21st, 2021 (until May 21st 2021). As discussed in detail in Ma (2020), the selection of starting points does not influence the correctness of prediction since the infection (or death) numbers before truncation points are accumulated and treated as the new starting infection (fatality) numbers for model-building.

Fig 2 displays the fitting of TPL model to the COVID-19 fatality datasets, and the TPL parameters are used to compute the confidence intervals for the fatality number prediction from PLEC-FTR model. Fig 3 displays the predicted COVID-19 fatalities based on the results listed in Table 1.

**Table 1**. The PLEC-FTR model (Power Law with Exponential Cutoff for Fatality-Time Relationship) model fitted with nonlinear optimization for daily cumulative counts of COVID-19 fatality, augmented with TPL to obtain the 95% confidence intervals

| Continent | $z$ | $d$ | $c$ | $R^2$ | $T_{max}$ | $T_{max}$ (Date) | $F_{max}$ | Observed (May 21) | Completion Level (%) | Lower Limit (95%) | Upper Limit (95%) |
|---|---|---|---|---|---|---|---|---|---|---|---|
| Africa | 1.150 | -0.002 | 180.452 | 1.000 | 501 | 3-Aug-2022 | 182,643 | 127,983 | 70.1 | 169,865 | 195,420 |
| Asia | 1.876 | 0.000 | 97.019 | 0.999 | NA | NA | NA | 636,068 | NA | NA | NA |
| Europe | 1.301 | -0.012 | 1734.846 | 1.000 | 113 | 11-Jul-2021 | 1,100,080 | 1,060,982 | 96.5 | 929,517 | 1,270,643 |
| North America | 1.185 | -0.009 | 983.515 | 0.999 | 129 | 28-Jul-2021 | 875,359 | 854,545 | 97.6 | 749,159 | 1,001,560 |
| South America | 1.323 | -0.007 | 1504.372 | 1.000 | 193 | 29-Sep-2021 | 952,175 | 762,185 | 80.0 | 839,676 | 1,064,675 |
| Oceania | 1.413 | -0.007 | 0.514 | 0.989 | 197 | 1-Oct-2021 | 1,191 | 1,095 | 92.0 | 1,075 | 1,306 |
| World# | 1.248 | -0.003 | 4957.140 | 1.000 | 485 | 19-Jul-2022 | 5,917,523 | 3,442,873 | 58.2 | 5,452,899 | 6,382,148 |

*Using fatality-time (date) data from March 21st to May 21st 2021.

**Table 2**. The PL-FTR model (Power Law for Fatality-Time Time Relationship) fitted for the daily cumulative counts of COVID-19 fatality, augmented with TPL to obtain the 95% confidence intervals

| Continent/ World | $z$ | $\ln(c)$ | $R$ | $P$-value | Observed (May 21) | Predicted (May 21) | Predicted (June 21) | Predicted (July 21) | Predicted (Aug 21) | Start date |
|---|---|---|---|---|---|---|---|---|---|---|
| **Asia** | 2.072 | 0.498 | 0.994 | 0.000 | 636,068 | 606,878 [562,269 651,487] | 687,070 [637,883 736,257] | 772,420 [718,484 826,356] | 862,949 [804,096 921,802] | 10-Feb-2020 |

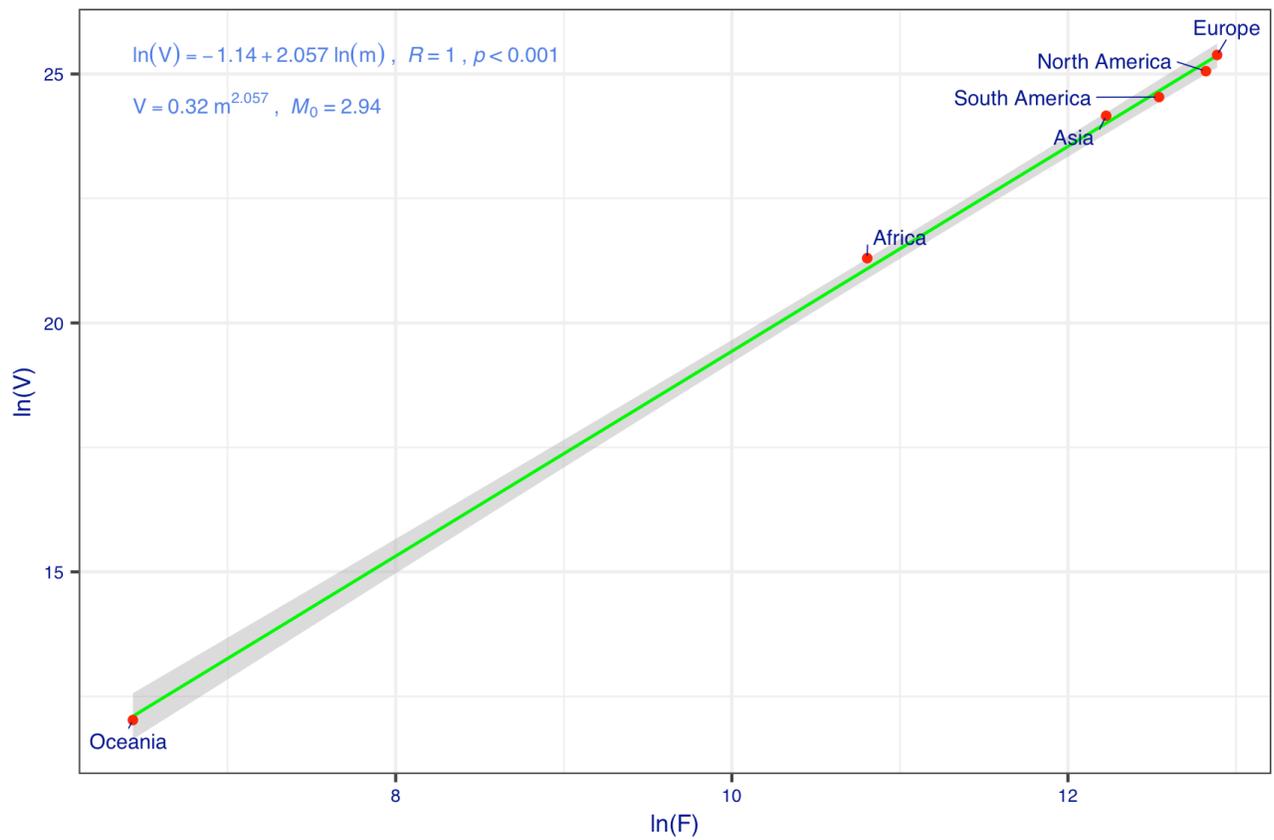

**Fig 2**. TPL (Taylor's power law) model fitted to the cumulative counts of COVID-19 fatalities: the variance corresponding to the fatality (F) is used to compute the standard error and width of confidence interval

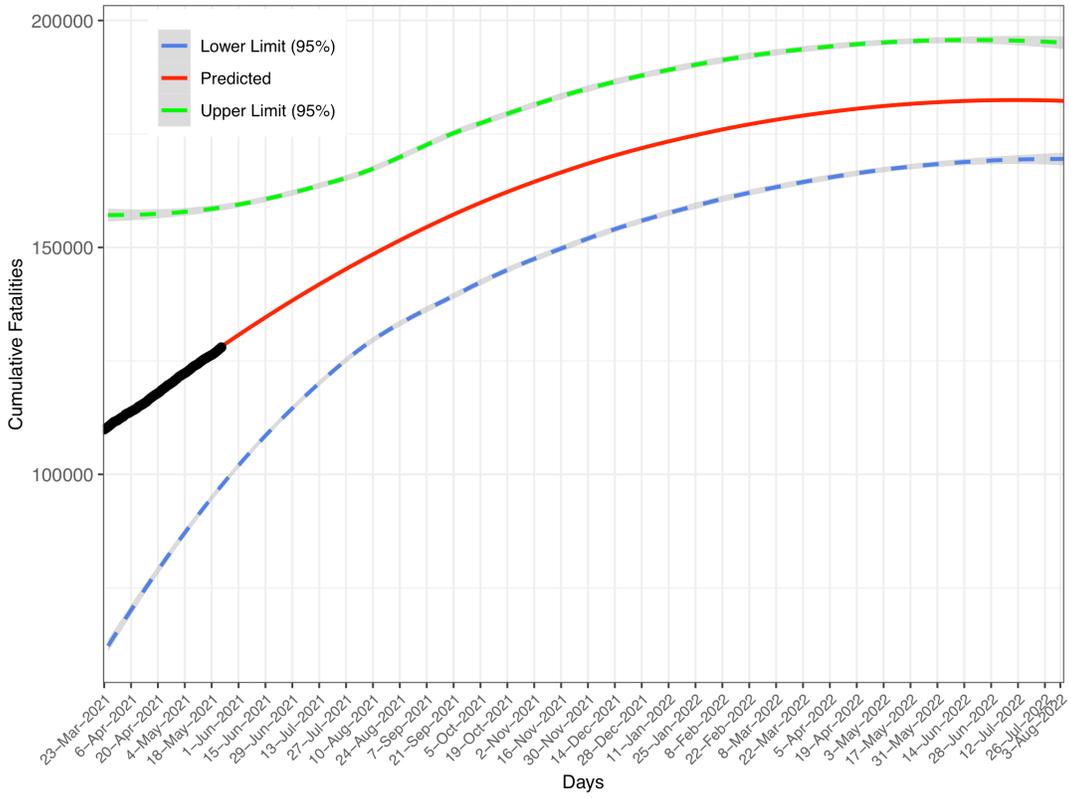

**Africa**

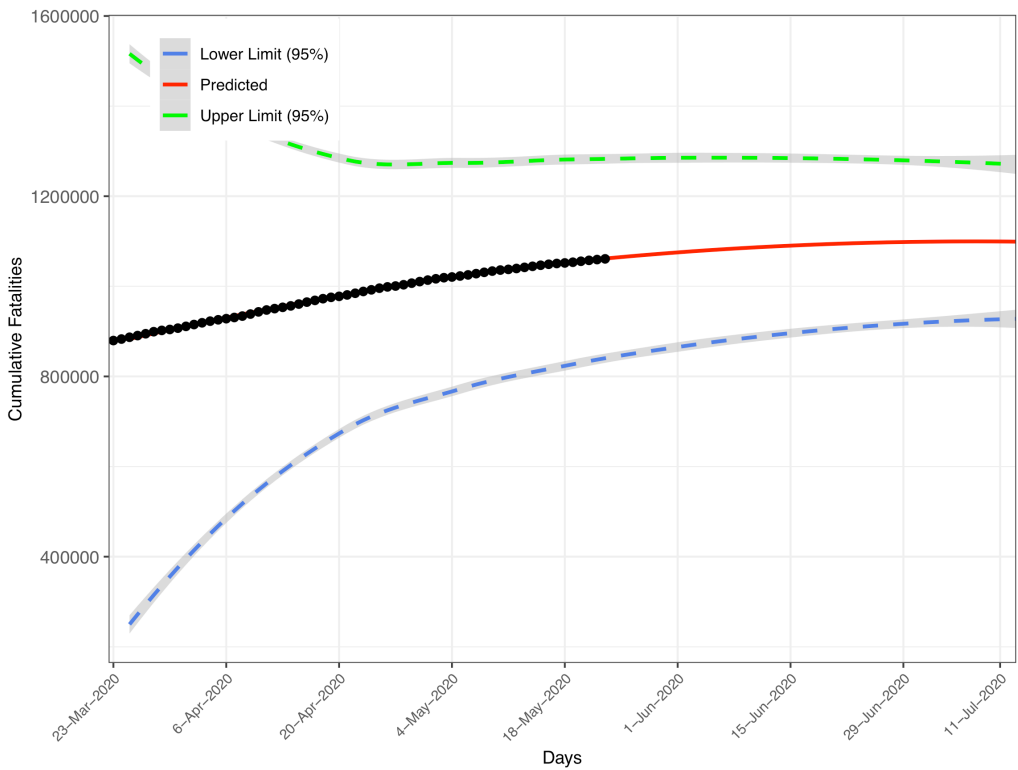

**Europe**

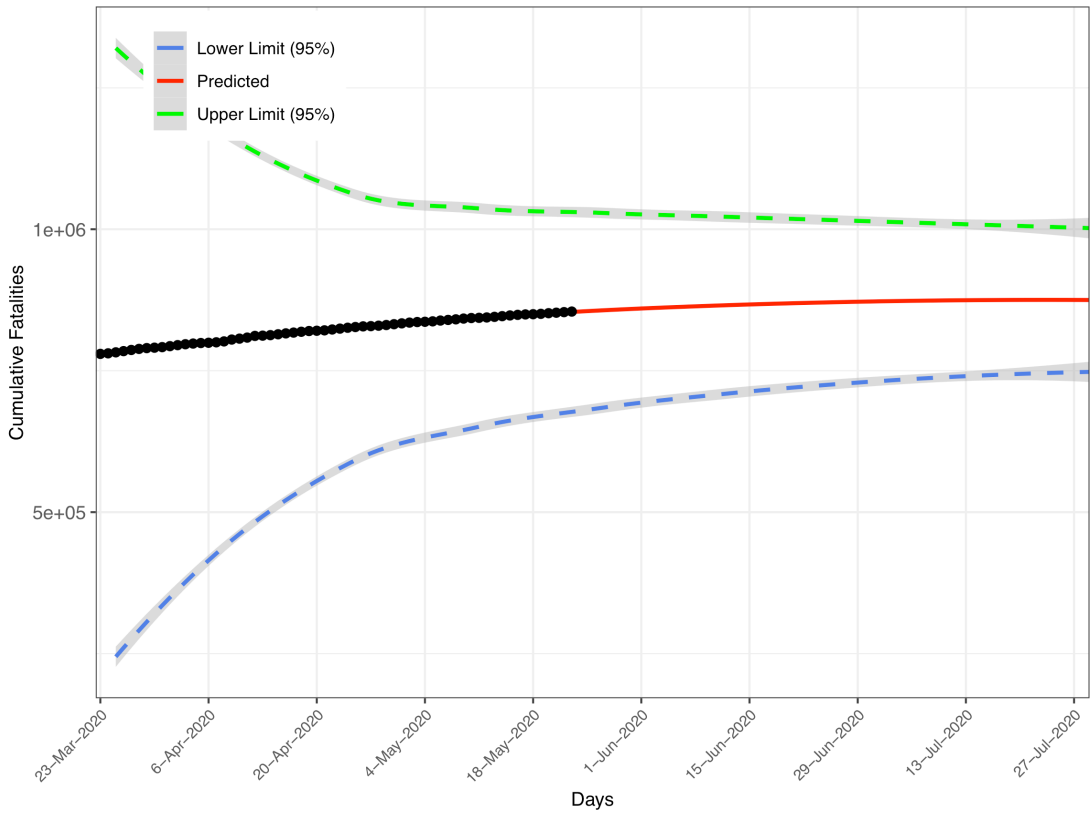

**North America**

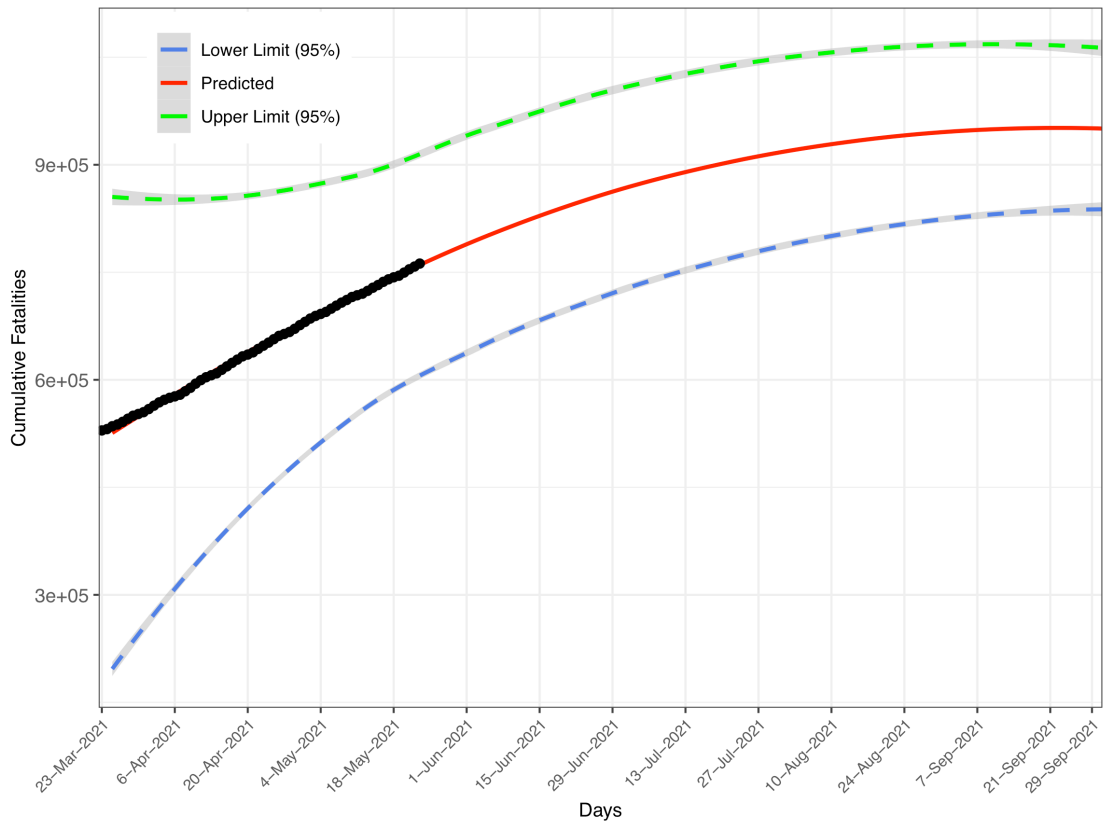

**South America**

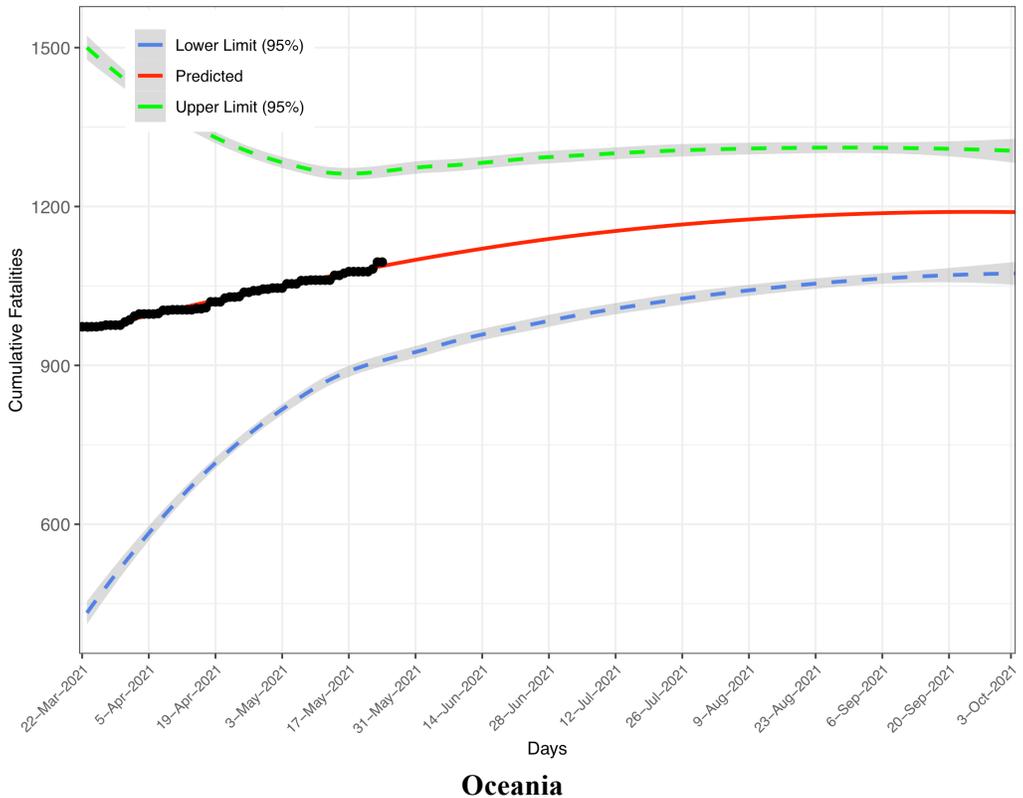

**Oceania**

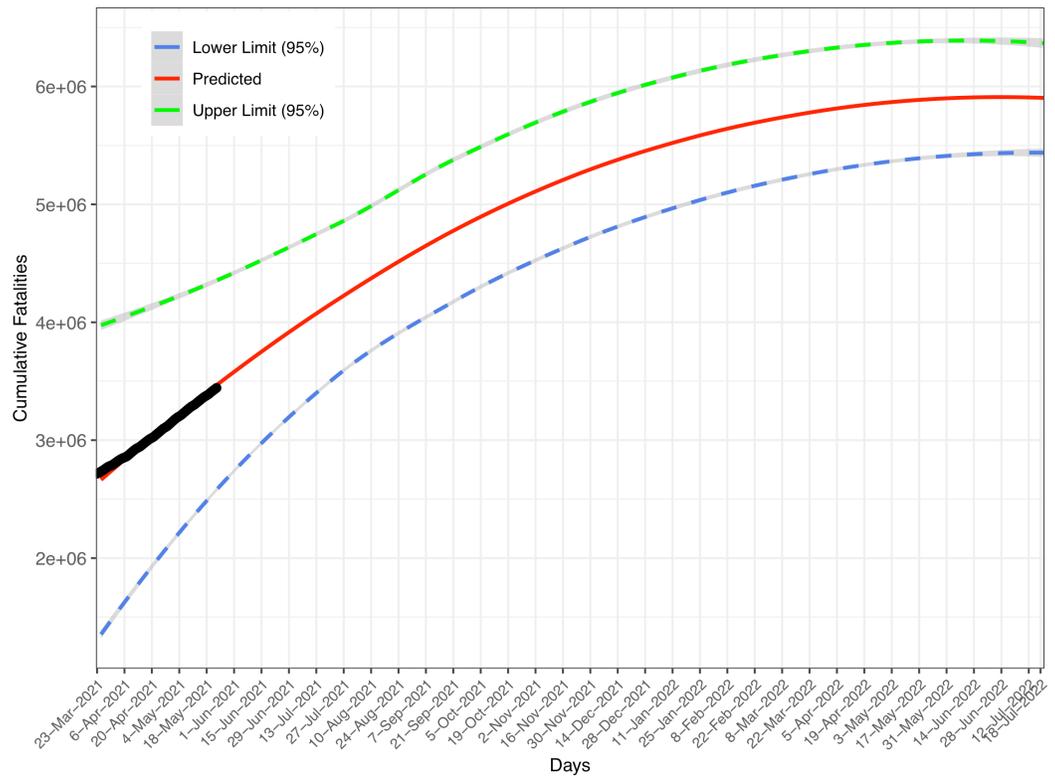

**World**

**Fig 3**. Predicted fatality number (solid curve), lower and upper bounds (dashed lines), and observed fatality number (solid cycles) for five continents and the world: Africa, Europe, North America, South America, Oceania, and World.

**Coupling TPL and PLEC-DAR for predicting the intervals of gut microbiome diversity**

Fig 1B showed the procedures for integrating the TPL and PLEC-DAR power law models for estimating the confidence intervals of biodiversity. Table 3 exhibited the results from implementing the coupled TPL and PLEC-DAR modeling analysis. The first five columns in Table 3 are simply the parameters of fitted PLEC-DAR model for the AGM (American gut microbiome) datasets, and the last four columns are simply the predicted maximal accrual diversity (species richness) of the AGM, including the $D_{max}$ (maximal accrual species richness), as well as the lower and upper limits of $D_{max}$. $A_{max}$ is the number of individuals (sample sizes) at which the $D_{max}$ is reached. Given that the samples of 1473 individuals are used to build the PLEC-DAR model, and the $A_{max}$ implies that 533 (2006-1473) additional individuals are required to accumulate the maximal accrual species richness in the AGM cohort or population. Fig 4 illustrated the fitting of TPL model, which helps the estimation of the 95%-level confidence intervals of $D_{max}$. Fig 5 illustrated the predicted species richness ($D_{max}$) (the solid curve in red color) and its confidence interval (dashed lines), and the observed species richness (the solid dots in blue color).

**Table 3**. The PLEC-DAR model (Power Law with Exponential Cutoff for Diversity-Area Relationship) fitted with 1000 times of re-sampling of the AGP datasets of 1473 American gut microbiome samples, augmented with TPL to obtain the 95% confidence intervals

| Dataset | $z$ | $d$ | $\ln(c)$ | $R$ | $A_{max}$ | $D_{max}$ | Lower Limit (95%) | Upper Limit (95%) |
|---|---|---|---|---|---|---|---|---|
| AGP Microbiome Species Richness | 0.386 | -0.0002 | 6.598 | 0.995 | 2006 | 9414 | 9310 | 9518 |

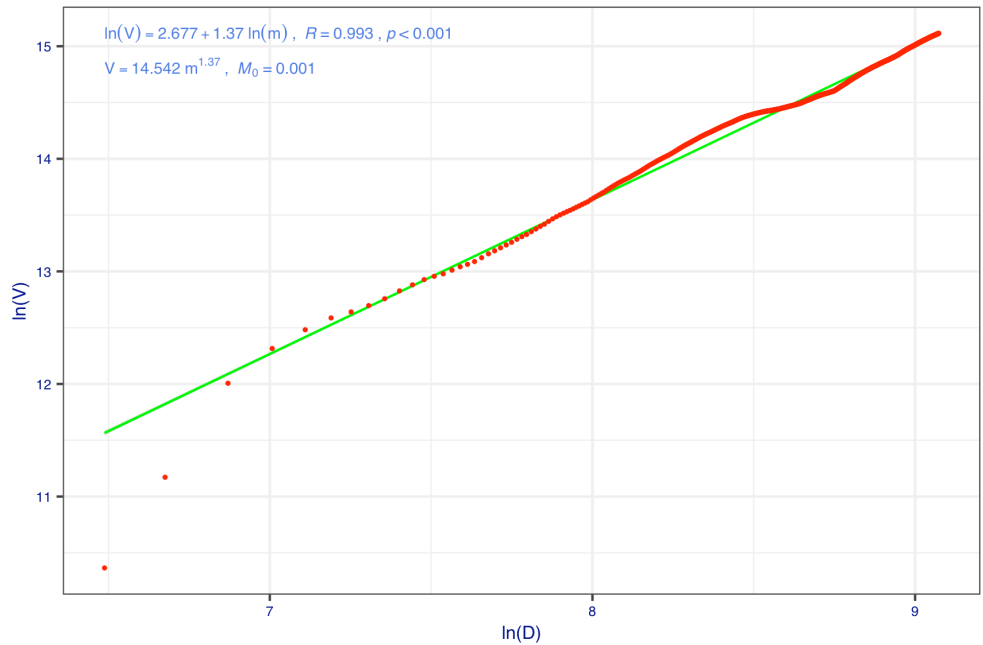

**Fig 4**. TPL (Taylor's power law) model fitted to the cumulative species richness of AGP data set (The 100 times of re-sampling were used to fit 100 PL-models, and here is one example; for each time of re-sampling, there are 1473 pairs of variance/mean of species richness, computed for each step of DAR accumulation).

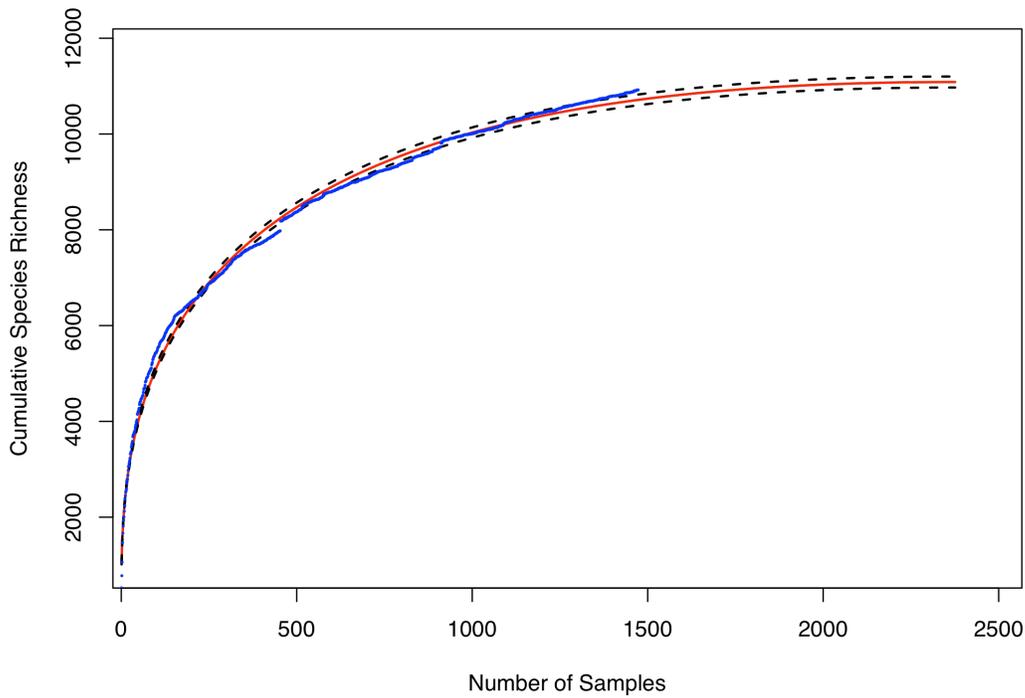

**Fig 5**. Predicted species richness (solid curve) of AGM (American gut microbiome) including lower and upper bounds (dashed lines) and observed species richness (solid cycles)

## Conclusions and Discussion

The following findings can be summarized from previous sections:

(*i*) Coupling of TPL (Taylor's power law) and PLEC (power-law with exponential cutoff) models, the two power laws from classic ecological theories with applications beyond their original domain of ecology and biology, offers a feasible solution for some important prediction problems. We demonstrate the approach with two examples.

(*ii*) For the COVID-19 prediction problem, the PLEC-FTR (fatality-time relationship) is able to predict the turning (inflection) points of fatality in the form of ($F_{max}$, $T_{max}$), *i.e.*, the maximal accrual fatality number and corresponding date at which $F_{max}$ is reached. In a previous study, we have demonstrated that the PLEC model successfully predicted the turning points of COVID-19 infections (Ma 2020). Both fatality and infection prediction problems are essentially the same, and therefore, prediction of fatality is undoubtedly feasible. An issue with our previous infection prediction is the lack of confidence interval. Thanks to the coupling with TPL model, the PLEC-FTR is able to deliver the confidence interval for $F_{max}$ by leveraging the capability of TPL in predicting variance (standard deviation) at different fatality level. This is because the TPL in the case of fatality prediction can be harnessed to establish the power-function relationship between mean fatality number and corresponding variance. With the variance (standard deviation), estimation of confidence intervals is then a trivial statistical exercise. Obviously, the coupling approach is equally applicable to the prediction of COVID-19 infections, although it was not proposed (Ma 2020). This example also suggests that the TPL-PLEC coupling approach may be applied to other similar predictive problems in epidemiology and public health.

(*iii*) For the biodiversity prediction of American gut microbiomes, the coupling of TPL and PLEC-DAR models are able to predict the maximal accrual species richness ($D_{max}$) of American gut microbiomes, which can be considered as potential or 'dark' species richness of gut microbiomes in the American cohort (population). The potential or dark biodiversity refers to the total diversity that includes the portion that may be absent locally but is present in regional species pool (and therefore is able to colonize local communities through dispersal/migration) (Ma 2019). In the case of human gut microbiome, the potential diversity can be considered as a cohort or population level characteristic of gut microbiome. In the case of this study, it can represent the potential species richness of the American population, given the datasets were obtained from sampling 1473 Americans, a sufficiently large sample size.

In perspective, we expect that the power-law coupling approach possesses great promises for a wide range of important problems whenever both TPL and PLEC models can be successfully applied. The precondition that both power law models must be reliably built also reminds us that the approach cannot be a silver-bullet solution. For example, in the case of PLEC-DAR modeling for the gut microbiome diversity, we only presented the results for species richness (*i.e.*, the Hill numbers when diversity order *q*=0). The reason was that TPL failed to fit the mean and variance of the Hill numbers at other diversity orders. This made it infeasible to estimate the confidence intervals for other diversity orders. TPL has been found applicable in many natural and man-made systems; however, there are situations it may fail. Taylor (2019) conjectured that TPL might work poorly for ratios and very poorly for bounded ratios. The Hill number at diversity *q*=0 (i.e., species richness) is integer, but at other diversity orders such as q=1, 2, or 3, the Hill numbers are indeed bounded ratios. Taylor (2019) conjecture may explain the limitation of TPL in fitting the mean-variance relationship in biodiversity measurements.

Finally, the *universality* property of power laws hints great promises for our coupling approach, although there have been occasional debates on proving universally in practical data fitted to power laws (*e.g.*, Stumpf & Porter 2012). The universality refers to the equivalence of power laws with a particular scaling parameter (exponent), such as *b* in TPL, *z* in SAR (DAR), or *w* in STR (DTR), which are termed *critical exponents*. Critical exponents are termed so because the power law distributions of certain quantities are associated with phase transitions in dynamic systems as they approach to criticality. The hallmark of universality is therefore the sharing of dynamics, and the systems with precisely the same critical exponents are said to belong to the same universality class. In the field of TPL, the transitions between aggregated (heterogeneous), random (Poisson), and uniform distribution of biological population or species abundance distribution can be characterized by the population aggregation critical density (PACD) (Ma 1991) or community heterogeneity critical diversity (CHCD) (Ma 2005), which could be generated by self-organizations in the ecosystems (*e.g.*, population or community). Different from physics, the processes such as self-organization in biology and ecology are difficult to prove vigorously. Nevertheless, there are indeed observations of the equivalence of TPL scaling exponents, such as the apparent invariance (constancy) of TPL scaling parameter (*b*) of global hot spring microbiomes across wide ranges of *pH*-values and temperatures (Li & Ma 2018). If these observations are found general in ecosystems, then the predictions based on our coupling approach of power laws can be not only feasible but also reliable. Unlike the events that are governed by the normal (Gaussian) distribution, the events governed by the highly skewed power law distribution are particularly challenging to predict. In particular, some power-law governed events are often lack of well-defined average value, but with potentially unbounded variance, tend to be black-swan and/or catastrophic; this also makes our proposed coupling method particularly valuable potentially.

## References (36 Citations)